\documentclass{Interspeech2024}
\usepackage{algorithm}
\usepackage{algpseudocode}
\usepackage{multirow}
\usepackage{arydshln}
\usepackage{hyperref,xurl}
\usepackage{graphicx}
\usepackage{float}
\usepackage{subcaption}
\usepackage{booktabs}
\usepackage{color}

\interspeechcameraready 

\title{Total-Duration-Aware Duration Modeling for Text-to-Speech Systems}

\name[]{Sefik Emre}{Eskimez}
\name[]{Xiaofei}{Wang}
\name[]{Manthan}{Thakker}
\name[]{Chung-Hsien}{Tsai}
\name[]{Canrun}{Li}
\name[]{Zhen}{Xiao}
\name[]{Hemin}{Yang}
\name[]{Zirun}{Zhu}
\name[]{Min}{Tang}
\name[]{Jinyu}{Li}
\name[]{Sheng}{Zhao}
\name[]{Naoyuki}{Kanda}

\address{
  Microsoft Corporation, USA}
\email{\{firstname.lastname\}@microsoft.com}


\begin{document}

\maketitle

\begin{abstract}
Accurate control of the total duration of generated speech by adjusting the speech rate is crucial for various text-to-speech (TTS) applications. However, the impact of adjusting the speech rate on speech quality, such as intelligibility and speaker characteristics, has been underexplored. In this work, we propose a novel total-duration-aware (TDA) duration model for TTS, where phoneme durations are predicted not only from the text input but also from an additional input of the total target duration. We also propose a MaskGIT-based duration model that enhances the diversity and quality of the predicted phoneme durations. Our results demonstrate that the proposed TDA duration models achieve better intelligibility and speaker similarity for various speech rate configurations compared to the baseline models. We also show that the proposed MaskGIT-based model can generate phoneme durations with higher quality and diversity compared to its regression or flow-matching counterparts.
\end{abstract}

\section{Introduction}

For many text-to-speech (TTS) applications, it is crucial that the total duration of the generated speech can be accurately adjusted to the target duration by modifying the speech rate. For example, in a video dubbing scenario, the output speech must match or closely approximate the duration of the source audio to ensure synchronization with the video~\cite{brannon2023dubbing}. The most naive solution is linearly scaling the speech rate to adhere to the target duration. However, it sometimes significantly degrades the speech quality such as intelligibility and speaker characteristics. In this paper, we investigate the duration model of TTS, focusing on achieving the desired target total duration (or target speech rate) that the TTS system should adhere to while preserving speech quality as much as possible.
 
There are two primary methods for duration modeling: implicit and explicit. In the case of implicit duration modeling, the TTS models do not have a separate mechanism to control phoneme duration, and the duration of each phoneme (or speech unit) is implicitly learned by the TTS model during the training~\cite{wang2023neural,wang2023speechx,gao2023e3}. In contrast, in the case of explicit duration modeling~\cite{beliaev2020talknet,kim2020glow,kim2021conditional,elias2021parallel,ren2021fastspeech,abbas2022expressive,effendi2022duration,jiang2023mega,le2023voicebox,vyas2023audiobox, mehta2023matcha}, a separate duration model estimates phoneme duration given the text input, and the audio model will generate audio conditioned by the estimated phoneme duration. 

In explicit duration modeling, a neural network is usually trained with mean-squared error (MSE) loss. The duration values are often converted to the logarithmic domain to make the prediction easier for the neural network~\cite{beliaev2020talknet,ren2021fastspeech,le2023voicebox,vyas2023audiobox,mehta2023matcha}. Recently, Le and Vyas et al.~\cite{le2023voicebox,vyas2023audiobox} proposed to train a duration model based on the masked duration prediction objective and showed better results in the zero-shot TTS setting. The authors also proposed a duration model based on flow-matching (FM), leading to better diversity for generated duration sequences that come with a cost of lesser intelligibility. 

A key advantage of explicit duration modeling is the ability to manipulate predicted durations before synthesizing the audio. It was demonstrated by various examples such as those in~\cite{ren2019fastspeech}. However, the impact of changing the duration on the audio quality has not been explored well. Effendi et al.~\cite{effendi2022duration} analyzed post-processing methods that normalize the predicted durations so that they sum up to the target total duration in the context of video dubbing systems. However, their analysis still lacks a detailed investigation into the impact of changing the speech rate. In our preliminary experiments, we found out that some of the metrics, such as intelligibility and speaker characteristics, degrade significantly, especially for faster speech.

In this paper, we present the Total-Duration-Aware (TDA) duration model that is designed to precisely control the length of generated speech while maintaining speech quality at different speech rates. This is achieved by incorporating the target total duration as an additional input, enabling the duration model to be aware of the target total duration and resulting in significantly improved intelligibility and preservation of speaker characteristics. We also introduce a novel duration model based on MaskGIT-based~\cite{chang2022maskgit} to enhance the diversity and quality of the phoneme durations. We evaluate our model in a zero-shot TTS setting with LibriSpeech~\cite{panayotov2015librispeech} and show the superiority of the proposed duration models across various speech rate configurations in both objective and subjective evaluations.

\begin{figure*}[t!]
\centering
\includegraphics[width=0.8\textwidth]{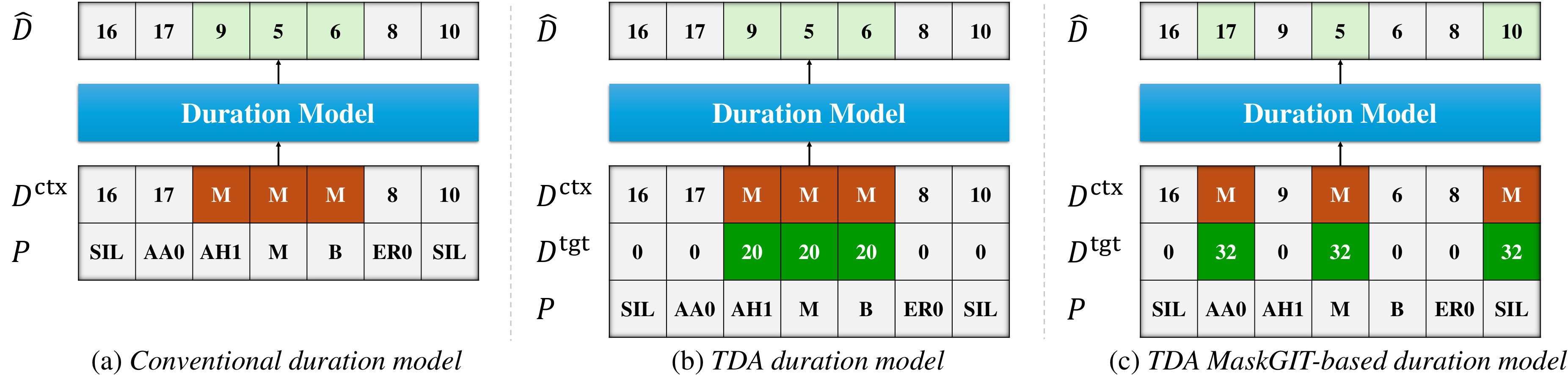}
\caption{Training inputs and targets for the conventional (baseline) and Total-Duration-Aware (TDA) models.}
\label{fig:training}
\vspace{-0.2in}
\end{figure*}

\section{Method}

We assume an input text for TTS is converted to a phoneme sequence $P=[p_1, p_2, ..., p_N]$, where $N$ is the length of the phoneme sequence, based on the grapheme to phoneme conversion.
The goal of the duration modeling is to estimate a duration sequence $\tilde{D}=[\tilde{d}_1, \tilde{d}_2, ..., \tilde{d}_N]$ for the phoneme sequence $P$. 
In this work, we further assume a duration context $D^{\rm ctx}=[d^{\rm ctx}_{1}, d^{\rm ctx}_{2}, ..., d^{\rm ctx}_{N}]$ and a target total duration $d^{\rm tgt}$ as additional inputs.
 The duration context $D^{\rm ctx}$ is a masked duration sequence~\cite{le2023voicebox}, where unknown durations are 0 and the known durations are equal to their corresponding phoneme durations (see Fig.~\ref{fig:training} (a)). It enables a duration estimation for zero-shot TTS and speech editing where a part of the duration sequence is known in inference~\cite{le2023voicebox}.
 On the other hand,  $d^{\rm tgt}$ constrains $\tilde{D}$ to be the target duration.
Let us define a mask sequence $M=[m_1, m_2, ..., m_N]$ where $m_n\in\{0,1\}$ represents if the $n$-th index is masked ($m_n=1$) or not ($m_n=0$) in $D^{\rm ctx}$. The estimated duration for the masked index must be equal to the target duration, expressed as $\sum_{n=1}^{N}m_n\tilde{d}_n=d^{\rm tgt}$.

\subsection{Baseline duration model}
As a baseline system, we use a neural network $G$ with parameters $\theta$ to predict a duration sequence $\hat{D}=[\hat{d}_1, \hat{d}_2, ..., \hat{d}_N]$ from the phoneme sequence and the duration context, i.e., $\hat{D} = G(P, D^{\rm ctx};\theta)$, as shown in Fig.~\ref{fig:training} (a). To adjust the predicted durations to the target total duration, we employ a length regulator (LR)~\cite{ren2019fastspeech} module during inference:
\begin{align}
 \tilde{D} = LR(\hat{D}, \alpha), \label{eq:lr}
\end{align}
\noindent where $LR$ is a function to linearly scale the duration with a scaling factor $\alpha$ computed as follows~\cite{effendi2022duration}:
\begin{align}
 \alpha = \frac{d^{\rm tgt}}{\sum_{n=1}^{N}m_n\hat{d_n}}.
\end{align}
\noindent 
Note that we only adjust the predicted durations for the masked indexes. 

The neural network $G$ can be modeled as a regression model trained by the MSE loss, or can be formulated as a flow-matching model~\cite{le2023voicebox} trained by conditional flow-matching objective~\cite{lipman2023flow}. We refer to the former as ``{\bf regression+LR}'' and the latter as ``{\bf FM+LR}''.

\subsection{TDA duration model}

To extend the baseline system, we propose to include the target total duration of the phonemes with unknown durations, $d^{\rm tgt}$, as an additional input to the network (Fig.~\ref{fig:training} (b)). This input allows the network to modify the generated durations as needed. We construct this input as follows: we make a new sequence, called the $D^{\rm tgt}=[m_1d^{\rm tgt}, m_2d^{\rm tgt}, ..., m_Nd^{\rm tgt}]$, that has the same length as the context input $D^{\rm ctx}$. For duration control models, the network is defined as $\hat{D} = G^{\rm TDA}(P, D^{\rm ctx}, D^{\rm tgt};\theta)$.

This approach adjusts the total length based on $d^{\rm tgt}$ but with limited accuracy. A post-processing step (Eq.~\eqref{eq:lr}) is still necessary to normalize durations and meet the target duration constraint. As with the baseline model, we can formulate the neural network $G^{\rm TDA}$ based on either the regression model or the flow-matching model. We refer the former as ``{\bf TDA regression+LR}'' and the latter as ``{\bf TDA FM+LR}''.

As a variant of the TDA duration model, we can integrate the normalization step into the network and train it end-to-end. We refer to the regression-based model with such an integrated normalization step as ``{\bf TDA regression+E2E}''. Note that it is not straightforward to train such an end-to-end model with FM because FM models a vector field rather than the output distribution. Therefore, we investigated only the ``TDA regression+E2E'' model in this paper.

\begin{algorithm}[H]
\scriptsize
\caption{Decoding algorithm for the \textit{TDA MaskGIT+LR} model. }\label{alg:maskgit}
\begin{algorithmic}
\State $G^{\rm TDA}$: Duration model
\State $P$: input phoneme sequence
\State $D^{\rm ctx}$: input masked duration sequence, only masked values will be predicted
\State $D^{\rm tgt}$: target duration sequence
\State $c$: confidence value
\State $\gamma$: scheduling function
\State $M^t$: mask sequence at step t, 1 for known, 0 for unknown values
\For{$t \gets 1$ to $T$}   
        \State {$D_{probs}$ $\gets$ {$G^{\rm TDA}(P, D^{\rm ctx}, D^{\rm tgt};\theta)$}} \Comment{Predict probabilities}
        \State {$\hat{D}$, $c$ $\gets$ {$Sample(D_{probs})$}} \\ \Comment{Sample duration values and their confidence scores}
        \State {$\hat{D}_{norm}$ $\gets$ {$UniformNormalize(\hat{D})$} } \\\Comment{Normalize the sampled durations to sum up to target duration}
        \State {$k$  $\gets$ $[\gamma(\frac{t}{T})N]$} \Comment{Get the number of tokens to fill for this iteration}
        \State {$M^{(t+1)}$ $\gets$ {$UnmaskTopN(c, k)$}} \Comment{Select the top n most confident frames}
        \State {$M_{diff}$ $\gets$ $abs(M^{(t+1)}-M^{(t)})$} \Comment{Identify filled tokens}
        \State {$D^{\rm ctx}[M_{diff}]$ $\gets$ $\hat{D}_{norm}[M_{diff}]$} \Comment{Update context}
        \State {$d^{\rm tgt}$ $\gets$ $d^{\rm tgt} - sum(\hat{D}_{norm}[M_{diff}])$}  \Comment{Update target duration}
        \State {$D^{\rm tgt}$ $\gets$ $[m_1^{(t+1)}d^{\rm tgt}, m_2^{(t+1)}d^{\rm tgt}, ..., m_N^{(t+1)}d^{\rm tgt}]$}
\EndFor
\end{algorithmic}
\end{algorithm}

\subsection{MaskGIT-based duration model}

MaskGIT~\cite{chang2022maskgit} is a bidirectional transformer decoder that can generate diverse and high-quality images with a constant number of iterations, $T$. In MaskGIT, a neural network is trained to predict the masked visual tokens, where the mask is applied in non-consecutive random regions. During inference, tokens are sampled for all masked regions, the top $k$ most confident tokens are selected, and the top $k$ tokens into the corresponding masked regions are filled. This procedure is iterated until all the masked region is filled. A scheduling function, $\gamma()$ controls $k$ for each iteration by adhering the number of iterations to be $T$.

In this work, we explore MaskGIT as the basis of high-quality duration modeling. First, we establish a baseline for the MaskGIT-based duration model. Unlike regression and FM models which estimate continuous representations as a duration for each phoneme, a discrete representation of durations, i.e., the number of frames of phonemes, is used for the MaskGIT-based duration model. During the training, the masked index, which is no longer contiguous like the baseline models, is randomly selected based on the cosine scheduling~\cite{chang2022maskgit}. The model is then trained based on cross-entropy loss to predict the discrete duration representation of the masked index. The inference is done similarly to the original MaskGIT except that we apply length normalization for each sampling iteration. Namely, after sampling the durations for all masked indices, we apply the LR module to these sampled durations so that the estimated duration follows $d^{\rm tgt}$. Then, we select the most confident $k$ tokens and fill them into their corresponding masked indices. The procedure is iterated until all the masked indices are filled with the estimated durations. We refer to this method as ``{\bf MaskGIT+LR}''.

Next, we incorporate the total duration input as an additional conditioning input to the network. Similar to the TDA duration models described in the previous section, we append $D^{\rm tgt}$ to the input sequence to condition the network by $d^{\rm tgt}$ (Fig.~\ref{fig:training} (c)). The training configuration remains the same as ``MaskGIT+LR''. During inference, we add another step to each iteration: we subtract the durations from the total target duration input, $d^{\rm tgt}$, at the end of the iteration. Consequently, in each iteration, $d^{\rm tgt}$ decreases until it reaches zero. We refer to this method as ``{\bf TDA MaskGIT+LR}''. The inference process is detailed in Algorithm~\ref{alg:maskgit}. 

\section{Experiments}

\begin{table}
\caption{Comparison of different models on sample diversity and quality for 1x rate, measured by the FDD metric, and the computational cost. All values were calculated with a 1x rate. We used a single generated sample for FM and MaskGIT models.}
\label{tab:fdd_results}
\vspace{-3mm}
\centering
\resizebox{\columnwidth}{!}{
\begin{tabular}{lrrl} 
\toprule
\textbf{model}    & \textbf{NFE or T} & \textbf{Cost} & \textbf{FDD$\downarrow$}  \\ 
\midrule
regression+LR (baseline)          & -                 & 1x            & $0.403_{\pm 0.001}$             \\
FM+LR             & 32                & 32x           & $0.318_{\pm 0.034}$                \\
MaskGIT+LR        & 32                & 32x           & \textbf{$0.206_{\pm 0.003}$}              \\           
\bottomrule
\end{tabular}
}
\vspace{-3mm}
\end{table}

To examine the effectiveness of our proposed duration models, we conducted objective and subjective evaluations. Specifically, we assessed the impact of changing the duration on the intelligibility as well as speaker similarity to the audio prompt in a setting of zero-shot TTS.  

\subsection{Training data and model configurations}
We trained our duration models based on LibriLight~\cite{kahn2020libri}. LibriLight consists of 60 thousand hours of unlabelled read speech in English from over 7,000 speakers. We used an off-the-shelf Kaldi-based automatic speech recognition (ASR) model\footnote{\tiny \url{https://kaldi-asr.org/models/m13}} trained on the 960-hour Librispeech data to transcribe the data, and used the frame-wise phoneme sequence from the ASR transcription for the duration model training. Our duration model closely followed the one presented in the Voicebox and Audiobox~\cite{le2023voicebox,vyas2023audiobox}. Specifically, we adopted a Transformer~\cite{vaswani2017attention} with the following specifications: 8 layers, 8 heads, 512 embedding dimensions, 2048 feed-forward network dimensions, and 1024 phone embedding dimensions. The layers had skip connections in the style of UNet. We used the same training settings as Audiobox~\cite{vyas2023audiobox}: masking all indexes probability of 0.2, random masking length from 10\% to 100\%, effective mini-batch size of 120K frames, and 600K training updates. For MaskGit models, we utilized a cosine scheduling function, to apply masking during training, with random uniform values between 0 to 1. We applied the log transform to the durations and $D^{\rm tgt}$ for both the network input and the loss calculations for the regression and FM models. For the MaskGIT-based duration model, we only applied the log transform to $D^{\rm tgt}$, and we set the maximum available duration value to 2048. 

Our audio model closely followed the Voicebox~\cite{le2023voicebox}. We used a Transformer with 24 layers, 16 heads, 1024 embedding dimensions, and 4096 feed-forward network dimensions. We pre-trained the audio model first on 200k-hour unlabeled anonymized data for 25.6M iterations using a modified version of SpeechFlow~\cite{liu2023generative} and fine-tuned it on LibriLight data for 640k iterations based on the conditional FM objective~\cite{wang2024ISsubmit}. 

\begin{figure}[t!]
\centering
\resizebox{\columnwidth}{!}{
\includegraphics[width=\columnwidth]{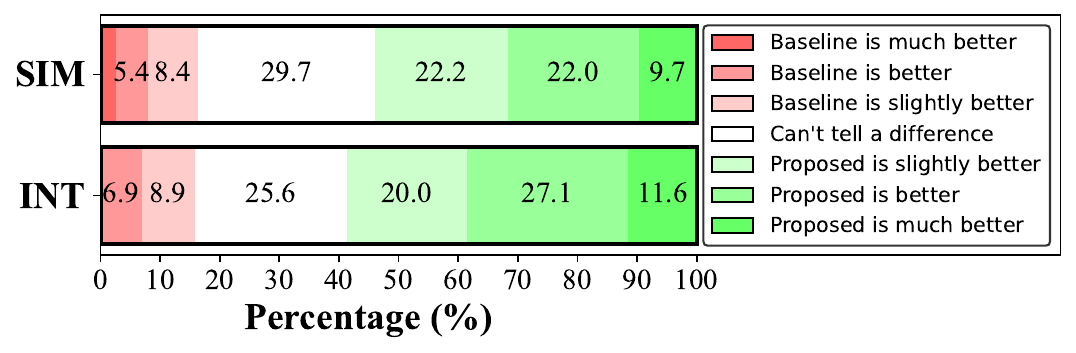}
}
\vspace{-7mm}
\caption{Subjective evaluation of intelligibility (INT) and speaker similarity (SIM) at 2x speech rate. }
\label{fig:subj_res}
\vspace{-0.2in}
\end{figure}

\begin{figure*}[t!]
\centering
\resizebox{\textwidth}{!}{
\includegraphics[width=\linewidth]{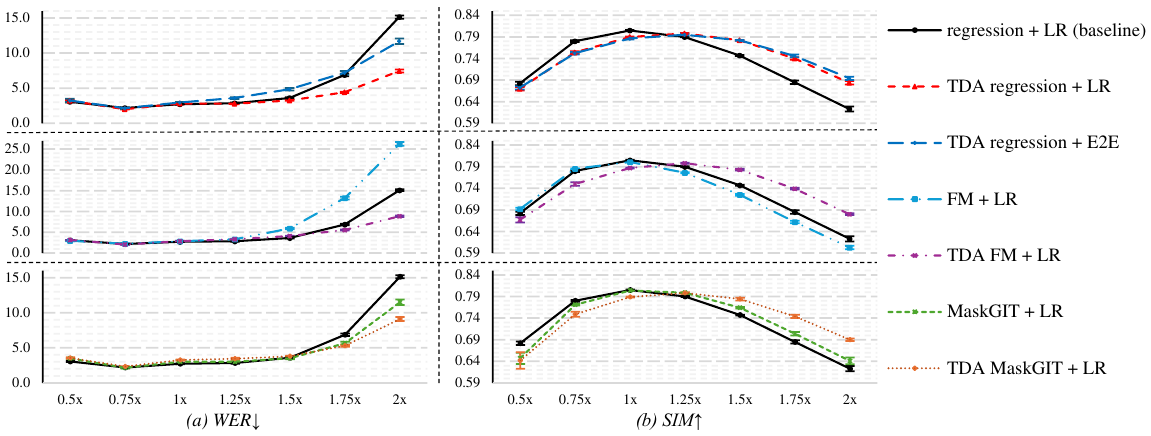}
}
\vspace{-0.2in}
\caption{WER (\%) and SIM for various speech rates, where the speech rate is computed by (the duration of the ground-truth speech) / (the target duration). The first, second, and third rows show the results of regression, FM, and MaskGIT models, respectively. 
}
\label{fig:main_results}
\vspace{-0.2in}
\end{figure*}

\subsection{Evaluation data, configuration, and metrics}
We evaluated our models on the ``test-clean'' of  LibriSpeech~\cite{panayotov2015librispeech} in a zero-shot TTS setting. By following~\cite{wang2023neural}, we filtered out the samples that were shorter than 4 seconds or longer than 10 seconds, resulting in 1237 samples. 

For each test sample, we used the first 3 seconds as the audio prompt and generated the entire audio, based on the transcription. Specifically, the transcription was first converted to the phoneme sequence based on PyKaldi toolkit\footnote{\tiny \url{https://github.com/pykaldi/pykaldi}}. Then, a duration model predicted the phoneme duration while being conditioned by the phoneme duration in the audio prompt as in~\cite{le2023voicebox}. Additionally, we appended 20 frames of silence to the predicted duration sequence before generating the audio. We applied classifier-free guidance (CFG) with a guidance strength of 0.7 and a number of function evaluations (NFE) of 32 for FM-based duration models. Following the approach of~\cite{le2023voicebox,vyas2023audiobox}, we used the mean of 8 duration samples as input to the audio model. For the MaskGIT duration models, $T$ was set to 32 to match the computational cost of FM models. The FM-based audio model was then applied to generate mel-spectrograms conditioned by the estimated duration as well as the audio prompt. The CFG with a guidance strength of 1.0 and NFE of 32 were used. Finally, the mel-spectrograms were converted to speech signals based on MelGAN-based vocoder~\cite{kumar2019melgan}.

We evaluated the generated speech objectively on three aspects: intelligibility, speaker similarity, and duration sample diversity and quality. For intelligibility, we used the word error rate (WER) where we averaged the WERs of three ASR models: HuBERT’s \textit{hubert-large-ls960-ft}\footnote{\tiny \url{https://huggingface.co/facebook/hubert-large-ls960-ft}}, NeMo’s \textit{stt\_en\_conformer\_transducer\_large}\footnote{\tiny \url{https://huggingface.co/nvidia/stt_en_conformer_transducer_xlarge}}, and Whisper’s \textit{whisper-medium}\footnote{\tiny \url{https://huggingface.co/openai/whisper-medium}}.  For speaker similarity (SIM), we extracted speaker embeddings using NeMo’s \textit{TitaNet-Large}\footnote{\tiny \url{https://huggingface.co/nvidia/speakerverification_en_titanet_large}} model and calculated the cosine distance between the original 3-sec audio prompt and the generated audio. For sample diversity and quality, we applied the Fréchet duration distance (FDD)~\cite{le2023voicebox} to measure the resemblance between the training and generated duration data distributions. To obtain reliable results, we generated the audio using three random seeds per experiment, computing the objective metrics for each seed, and reporting the average metrics as well as their standard deviation. In addition to these objective metrics, we also conducted subjective evaluations to measure intelligibility and speaker similarity, the details of which are described in the next section.

\subsection{Results}

\noindent\textbf{Regression vs. FM vs. MaskGIT:}
Before conducting the evaluation with various target durations, we evaluated the various duration models based on the FDD, computed across all non-silence phonemes. The result is shown in Table~\ref{tab:fdd_results}. In this experiment, we set the target duration as the same as that of the ground-truth speech. The ``MaskGIT+LR'' model scored significantly better FDD than the ``regression+LR'' and ``FM+LR'' models, confirming that MaskGIT can generate a closer duration distribution to the training data. 

\noindent\textbf{Baselines vs. TDA models:}
The result of the objective evaluation is presented in Fig.~\ref{fig:main_results}. In this experiment, we controlled the target duration based on the ground-truth duration, and the result is reported based on the speech rate, which is the duration of the ground-truth speech divided by the target duration. We started by analyzing the ``regression+LR'' which we refer to as the baseline model: SIM deteriorated when the rate deviated from 1x. This implies that the baseline model failed to preserve the speaker characteristics when the speech rate varies. Moreover, the 2x results exhibited a sharp increase in WER (from 2.7 to 15.1), implying a significant degradation of intelligibility. The proposed ``TDA regression+LR'' model surpassed the baseline model for the faster rates both in WER and SIM while maintaining comparable WER and SIM for the slower rates. A paired t-test was conducted for 2x rate results, showing a significant difference with p $<$ 0.05. In our experiment, the ``TDA regression+E2E'' model did not provide any notable benefit over the ``TDA regression+LR'' model, except for a minor SIM improvement at faster rates with a significant WER degradation.

We proceeded to examine the FM and MaskGIT-based models. The FM and MaskGIT models also gain from the additional $D^{\rm tgt}$ input, particularly for the higher speech rates. ``TDA FM+LR'' achieved considerable improvement over ``FM+LR'' for the 2x rate (WER: 26.2 vs. 8.9, SIM: 0.603 vs 0.680). Notably, ``FM+LR'' performed decently for the slower speech rates, but it is the worst performer among all models for the faster speech rates. Likewise, ``TDA MaskGIT+LR'' enhanced the results significantly over the ``MaskGIT+LR'' for the 2x rate (WER: 11.5 vs. 9.1, SIM: 0.641 vs 0.690). Another observation is that ``MaskGIT+LR'' delivered better SIM and WER for the 2x rate than the ``FM+LR''. 

The higher speech rates remain a challenge for improvement. The best WER result we obtained for the 2x model was 7.4 using ``TDA regression+LR'', but it was considerably worse than the 1x rate WER of 2.7. Likewise, the SIM improvement was modest for the 2x rate and far from the SIM value of 0.805 at the 1x rate. Moreover, for the lower speech rates, SIM slightly worsened. We note that we achieved these results using only 1x rate training data for both duration and audio models. Enriching the training data with diverse speech rates would produce better results, which we defer to future work. 

\noindent\textbf{Subjective evaluations:} We conducted a subjective evaluation of intelligibility and speaker similarity using 30 random samples at a 2x rate with 15 native English speakers. For the intelligibility test, we presented the subjects with the ``regression+LR'' (=baseline model) and ``TDA regression+LR'' (=proposed model) outputs, asking them to compare the samples and report a score between -3 (the proposed model is much worse than the baseline) to 3 (the proposed model is much better than the baseline). The order of the two samples was randomized so that the subjects could not know which sample was from which model. For the speaker similarity test, we presented the 3-sec audio prompt and the outputs from two models. Similar to the intelligibility test, we asked the listeners to compare the samples and report a score between -3 to 3 based on the similarity of the speaker compared to the audio prompt.  

The results are shown in Fig.~\ref{fig:subj_res}. In the intelligibility test, the subjects preferred ``TDA regression+LR'' over ``regression+LR'' for 58.7\% of the samples, while they could not tell a difference for 25.6\% of the samples. In the speaker similarity test, the subjects preferred ``TDA regression+LR'' over ``regression+LR'' for 53.9\% of the samples, while they could not tell a difference for 29.7\% of the samples. Overall, these results suggest that the proposed TDA model performed better in terms of both intelligibility and speaker similarity, as perceived by the subjects.

\section{Conclusions}
In this paper, we proposed a TDA duration model that predicts phoneme duration with additional conditioning on the total target duration. We also proposed a MaskGIT-based duration model that enhances the diversity and quality of the predicted durations. Our evaluation revealed that the proposed TDA duration model achieved significantly better intelligibility and speaker similarity especially when generating a fast-paced speech. In addition, we showed that our MaskGIT-based duration model can generate the phoneme duration sequence with significantly higher quality and diversity compared to the regression and flow-matching counterparts.

\bibliographystyle{IEEEtran}
\bibliography{mybib}

\begin{thebibliography}{10}
\providecommand{\url}[1]{#1}
\csname url@samestyle\endcsname
\providecommand{\newblock}{\relax}
\providecommand{\bibinfo}[2]{#2}
\providecommand{\BIBentrySTDinterwordspacing}{\spaceskip=0pt\relax}
\providecommand{\BIBentryALTinterwordstretchfactor}{4}
\providecommand{\BIBentryALTinterwordspacing}{\spaceskip=\fontdimen2\font plus
\BIBentryALTinterwordstretchfactor\fontdimen3\font minus \fontdimen4\font\relax}
\providecommand{\BIBforeignlanguage}[2]{{%
\expandafter\ifx\csname l@#1\endcsname\relax
\typeout{** WARNING: IEEEtran.bst: No hyphenation pattern has been}%
\typeout{** loaded for the language `#1'. Using the pattern for}%
\typeout{** the default language instead.}%
\else
\language=\csname l@#1\endcsname
\fi
#2}}
\providecommand{\BIBdecl}{\relax}
\BIBdecl

\bibitem{brannon2023dubbing}
W.~Brannon, Y.~Virkar, and B.~Thompson, ``Dubbing in practice: A large scale study of human localization with insights for automatic dubbing,'' \emph{Transactions of the Association for Computational Linguistics}, vol.~11, pp. 419--435, 2023.

\bibitem{wang2023neural}
C.~Wang, S.~Chen, Y.~Wu, Z.~Zhang, L.~Zhou, S.~Liu, Z.~Chen, Y.~Liu, H.~Wang, J.~Li \emph{et~al.}, ``Neural codec language models are zero-shot text to speech synthesizers,'' \emph{arXiv preprint arXiv:2301.02111}, 2023.

\bibitem{wang2023speechx}
X.~Wang, M.~Thakker, Z.~Chen, N.~Kanda, S.~E. Eskimez, S.~Chen, M.~Tang, S.~Liu, J.~Li, and T.~Yoshioka, ``{SpeechX}: Neural codec language model as a versatile speech transformer,'' \emph{arXiv preprint arXiv:2308.06873}, 2023.

\bibitem{gao2023e3}
Y.~Gao, N.~Morioka, Y.~Zhang, and N.~Chen, ``{E3 TTS}: Easy end-to-end diffusion-based text to speech,'' in \emph{Proc. ASRU}.\hskip 1em plus 0.5em minus 0.4em\relax IEEE, 2023, pp. 1--8.

\bibitem{beliaev2020talknet}
S.~Beliaev, Y.~Rebryk, and B.~Ginsburg, ``{TalkNet}: Fully-convolutional non-autoregressive speech synthesis model,'' \emph{arXiv preprint arXiv:2005.05514}, 2020.

\bibitem{kim2020glow}
J.~Kim, S.~Kim, J.~Kong, and S.~Yoon, ``{Glow-TTS}: A generative flow for text-to-speech via monotonic alignment search,'' in \emph{Proc. NeurIPS}, vol.~33, 2020, pp. 8067--8077.

\bibitem{kim2021conditional}
J.~Kim, J.~Kong, and J.~Son, ``Conditional variational autoencoder with adversarial learning for end-to-end text-to-speech,'' in \emph{Proc. ICML}.\hskip 1em plus 0.5em minus 0.4em\relax PMLR, 2021, pp. 5530--5540.

\bibitem{elias2021parallel}
I.~Elias, H.~Zen, J.~Shen, Y.~Zhang, Y.~Jia, R.~J. Weiss, and Y.~Wu, ``{Parallel Tacotron}: Non-autoregressive and controllable {TTS},'' in \emph{Proc. ICASSP}.\hskip 1em plus 0.5em minus 0.4em\relax IEEE, 2021, pp. 5709--5713.

\bibitem{ren2021fastspeech}
Y.~Ren, C.~Hu, X.~Tan, T.~Qin, S.~Zhao, Z.~Zhao, and T.-Y. Liu, ``Fastspeech 2: Fast and high-quality end-to-end text to speech,'' in \emph{Proc. ICLR}, 2021.

\bibitem{abbas2022expressive}
S.~{Ammar Abbas}, T.~Merritt, A.~Moinet, S.~Karlapati, E.~Muszynska, S.~Slangen, E.~Gatti, and T.~Drugman, ``Expressive, variable, and controllable duration modelling in {TTS},'' in \emph{Proc. Interspeech}, 2022, pp. 4546--4550.

\bibitem{effendi2022duration}
J.~Effendi, Y.~Virkar, R.~Barra-Chicote, and M.~Federico, ``Duration modeling of neural {TTS} for automatic dubbing,'' in \emph{Proc. ICASSP}.\hskip 1em plus 0.5em minus 0.4em\relax IEEE, 2022, pp. 8037--8041.

\bibitem{jiang2023mega}
Z.~Jiang, Y.~Ren, Z.~Ye, J.~Liu, C.~Zhang, Q.~Yang, S.~Ji, R.~Huang, C.~Wang, X.~Yin \emph{et~al.}, ``{Mega-TTS}: Zero-shot text-to-speech at scale with intrinsic inductive bias,'' \emph{arXiv preprint arXiv:2306.03509}, 2023.

\bibitem{le2023voicebox}
M.~Le, A.~Vyas, B.~Shi, B.~Karrer, L.~Sari, R.~Moritz, M.~Williamson, V.~Manohar, Y.~Adi, J.~Mahadeokar \emph{et~al.}, ``Voicebox: Text-guided multilingual universal speech generation at scale,'' \emph{arXiv preprint arXiv:2306.15687}, 2023.

\bibitem{vyas2023audiobox}
A.~Vyas, B.~Shi, M.~Le, A.~Tjandra, Y.-C. Wu, B.~Guo, J.~Zhang, X.~Zhang, R.~Adkins, W.~Ngan \emph{et~al.}, ``Audiobox: Unified audio generation with natural language prompts,'' \emph{arXiv preprint arXiv:2312.15821}, 2023.

\bibitem{mehta2023matcha}
S.~Mehta, R.~Tu, J.~Beskow, {\'E}.~Sz{\'e}kely, and G.~E. Henter, ``{Matcha-TTS}: A fast {TTS} architecture with conditional flow matching,'' \emph{arXiv preprint arXiv:2309.03199}, 2023.

\bibitem{ren2019fastspeech}
Y.~Ren, Y.~Ruan, X.~Tan, T.~Qin, S.~Zhao, Z.~Zhao, and T.-Y. Liu, ``Fastspeech: Fast, robust and controllable text to speech,'' in \emph{Proc. NeurIPS}, vol.~32, 2019.

\bibitem{chang2022maskgit}
H.~Chang, H.~Zhang, L.~Jiang, C.~Liu, and W.~T. Freeman, ``{MaskGIT}: Masked generative image transformer,'' in \emph{Proc. CVPR}, 2022, pp. 11\,315--11\,325.

\bibitem{panayotov2015librispeech}
V.~Panayotov, G.~Chen, D.~Povey, and S.~Khudanpur, ``{LibriSpeech: An ASR corpus based on public domain audio books},'' in \emph{Proc. ICASSP}.\hskip 1em plus 0.5em minus 0.4em\relax IEEE, 2015, pp. 5206--5210.

\bibitem{lipman2023flow}
Y.~Lipman, R.~T.~Q. Chen, H.~Ben-Hamu, M.~Nickel, and M.~Le, ``Flow matching for generative modeling,'' in \emph{Proc. ICLR}, 2023.

\bibitem{kahn2020libri}
J.~Kahn, M.~Rivi{\`e}re, W.~Zheng, E.~Kharitonov, Q.~Xu, P.-E. Mazar{\'e}, J.~Karadayi, V.~Liptchinsky, R.~Collobert, C.~Fuegen \emph{et~al.}, ``Libri-light: A benchmark for {ASR} with limited or no supervision,'' in \emph{Proc. ICASSP}.\hskip 1em plus 0.5em minus 0.4em\relax IEEE, 2020, pp. 7669--7673.

\bibitem{vaswani2017attention}
A.~Vaswani, N.~Shazeer, N.~Parmar, J.~Uszkoreit, L.~Jones, A.~N. Gomez, {\L}.~Kaiser, and I.~Polosukhin, ``Attention is all you need,'' in \emph{Proc. NeurIPS}, vol.~30, 2017.

\bibitem{liu2023generative}
A.~H. Liu, M.~Le, A.~Vyas, B.~Shi, A.~Tjandra, and W.-N. Hsu, ``Generative pre-training for speech with flow matching,'' in \emph{Proc. ICLR}, 2023.

\bibitem{wang2024ISsubmit}
X.~Wang, S.~E. Eskimez, M.~Thakker, H.~Yang, Z.~Zhu, M.~Tang, Y.~Xia, J.~Li, S.~Zhao, J.~Li, and N.~Kanda, ``An investigation of noise robustness for flow-matching-based zero-shot {TTS},'' in \emph{Proc. Interspeech}, 2024.

\bibitem{kumar2019melgan}
K.~Kumar, R.~Kumar, T.~De~Boissiere, L.~Gestin, W.~Z. Teoh, J.~Sotelo, A.~De~Brebisson, Y.~Bengio, and A.~C. Courville, ``{MelGAN}: Generative adversarial networks for conditional waveform synthesis,'' in \emph{Proc. NeurIPS}, vol.~32, 2019.

\end{thebibliography}

\end{document}